\documentclass[12pt,preprint]{aastex}

\newcommand{\um}{$\mu$m}
\newcommand{\z}{{\it z}}
\newcommand{\nwu}{nW~m$^{-2}$~sr$^{-1}$}
\newcommand{\ab}{$\sim$}
\newcommand{\sst}{{\it Spitzer}}

\slugcomment{To Appear in the {\it Astrophysical Journal Supplement Series}}

\shorttitle{Missing Link between ISOCAM and SCUBA}
\shortauthors{Chary et al.}

\begin{document}


\title{The Nature of Faint 24\um\ sources Seen in \sst\ Observations of ELAIS-N1}

\author{R. Chary\altaffilmark{1}, S. Casertano\altaffilmark{2}, M. E. Dickinson\altaffilmark{2,3}, H. C. Ferguson\altaffilmark{2}, P. R. M. Eisenhardt\altaffilmark{4}, D. Elbaz\altaffilmark{5}, N. A. Grogin\altaffilmark{6}, L. A. Moustakas\altaffilmark{2}, W. T. Reach\altaffilmark{1}, H. Yan\altaffilmark{1}} 

\altaffiltext{1}{MS220-6 Spitzer Science Center, Caltech, Pasadena CA 91125; rchary@caltech.edu}
\altaffiltext{2}{Space Telescope Science Institute, 3700 San Martin Dr., Baltimore, MD 21218}
\altaffiltext{3}{NOAO, 950 N. Cherry St., Tucson, AZ 85719}
\altaffiltext{4}{MS169-327, Jet Propulsion Laboratory, 4800 Oak Grove Dr., Pasadena, CA 91109}
\altaffiltext{5}{DAPNIA/Service d' AStrophysique, CEA-Saclay, 91191 Gif-sur-Yvette Cedex, France}
\altaffiltext{6}{Dept. of Physics \& Astronomy, Johns Hopkins University, 3400 N. Charles St., Baltimore, MD 21218}
 
\begin{abstract}
The \sst\ Space Telescope has undertaken the deepest ever observations of the 24\um\ sky in the ELAIS-N1 field with the MIPS instrument
as part of GOODS Science Verification observations. 
We present completeness corrected extragalactic source counts down to 
24$\mu$m flux densities of 
20$\mu$Jy (30\% completeness), a factor of 10000 more sensitive than IRAS. 
The shape of the counts confirms model predictions for a
strong evolution of the infrared luminosity function between redshift of 0 and 1,
and suggests a flattening in the evolutionary parameters at higher redshifts.
Models that fit the counts indicate that luminous
infrared galaxies (10$^{11}<$L(8-1000\um)$<10^{12}$~L$_{\sun}$) in the redshift
range 0.5$<z<2.5$ constitute $>$60\% of 
24\um\ sources seen in the flux range 20$<$S$_{\nu}<1000$~$\mu$Jy. 
At least 85\%, and possibly all of the 24\um\ sources have a counterpart in the IRAC 3.6\um\ and 4.5\um\ near-infrared channels
consistent with the expected
spectral energy distribution of infrared luminous galaxies at moderate redshift. The similarity between the observed
mid-infrared to near-infrared
flux ratios of the \sst\ detected sources and the 15$\mu$m/HK-band
flux ratios of the ISOCAM 15\um\ sources seen in the HDF-N strongly
suggests that faint 24\um\ sources are high-redshift analogs of ISOCAM 15\um\ sources, and that they have
the potential to provide an evolutionary
connection between the well-studied $z\sim3$ Lyman-break galaxy population and the dusty starburst galaxies seen at $z\sim1$.
\end{abstract}


\dataset{ads/sa.spitzer\#0006006016}
\dataset{ads/sa.spitzer\#0006006272}
\dataset{ads/sa.spitzer\#0006006528}

\section{Introduction}

Deep extragalactic surveys at mid-infrared and submillimeter wavelengths have revealed a population of
dusty galaxies with large mid- and far-infrared luminosities at intermediate and high 
redshifts \citep[see e.g.][and references therein]{fran02}. 
The number counts of these objects is more than an order of magnitude higher than expected if the local infrared
luminosity function of galaxies were constant with redshift. 
Model fits to the observed counts at these wavelengths by various groups suggest a rapid increase in the co-moving
number density of infrared luminous galaxies between redshifts of 0 and 1$-$1.5 followed by a flattening or decrease out to
higher redshifts
\citep[][; hereafter XU, CE, LDP and KRR respectively]{xu01, ce01, lag03, krr03}. 
The mid-infrared bandpasses i.e. ISOCAM 15\um\ and \sst\ 24\um\ filters trace rest-frame 
7.7\um\ and 12\um\ at \z\ab 1 and \z\ab2 respectively where the radiation is dominated
by the line and continuum emission from polycyclic aromatic hydrocarbons and very small
dust grains \citep[see e.g.][for a review]{dwek97}. 
As a result, it has been unclear if the evolutionary turn over at \z\ab 1 is because of the
$k$-correction in the ISOCAM 15\um\ filter which would selectively detect objects with strong 7.7\um\ PAH features
at that redshift
or because of a real change in the evolutionary parameters. 
In this paper, we present faint 24\um\ counts from \sst\ observations
of the European Large Area ISO Survey-N1 (ELAIS-N1) field,  
evaluate their contribution to the extragalactic background light and
compare the properties of these objects with the ISOCAM detected 15\um\ sources
with particular emphasis on their near-infrared/IRAC counterparts. Throughout this paper, we refer to luminous infrared
galaxies (LIGs) as objects with 10$^{11}<$L$_{{\rm IR}}<10^{12}$~L$_{\sun}$ and ultraluminous infrared galaxies (ULIGs) as
objects with L$_{{\rm IR}}>10^{12}$~L$_{\sun}$. 

\section{Observations}

The \sst\ observations of ELAIS-N1 ($\alpha=16h09m20s$; $\delta=54\arcdeg57\arcmin00\arcsec$; J2000) were undertaken 
under Director's Discretionary Time
as part of the Great Observatories Origins Deep Survey (GOODS) Science Verification program to assess the effect
of source confusion on ultradeep \sst\ surveys. The maximum integration time amounted to 4620sec per detector pixel of MIPS 24\um\
integration and 6900sec of IRAC observations corresponding to background limited 5$\sigma$ sensitivities of 6$\mu$Jy/pix
and 0.4$\mu$Jy/pix (3.6 \& 4.5\um)
respectively \citep[see][for instrument description]{faz04, rie04}. The output of the Spitzer
Science Center v9.5 pipeline was post-processed after making corrections for varying sky background, striping (jailbar) correction for
every fourth column due
to a bias drift in the readout electronics, 
and distortion corrections based on a grid of 2MASS stars and cross-channel
counterparts in the IRAC and MIPS images. The images were then
resampled (drizzled; Fruchter \& Hook 2002) 
onto a 1.2$\arcsec$ pixel grid for MIPS and 0.6$\arcsec$ for IRAC and sources extracted. 
The observations covered a total area of $\sim$165$\sq\arcmin$
with an integration time greater than 900s.
Two rectangular strips with a combined area of 50$\sq\arcmin$ have the deepest exposures ($>$4200sec). Figure 1 shows
a fraction of the 24\um\ image with dots representing the IRAC counterparts of the MIPS sources.

For the IRAC data, roughly half the field was covered by the 3.6\um\ and 5.8\um\ channels, while the other half was covered
by the 4.5 and 8.0$\mu$m channels. Only the 3.6 and 4.5$\mu$m IRAC images are considered here, which together provide the best
angular resolution and faintest flux limits currently available for studying the counterparts to the 24$\mu$m sources.
The source density in the IRAC channels was found to be about a factor of 4 higher than in the 24\um\ image at these depths.

Sources catalogs were generated using SExtractor \citep{ber96}. 
SExtractor parameters were fine tuned from simulated pure-noise images. These images were simulated
by generating noise-only data frames where the rms for each 30s frame
was the same as measured in the data. Each frame was then run through the post-processing pipeline,
applying the same distortion corrections and drizzle parameters in generating a final stack as for the real data. 
In general, we find that the majority of the sources in the real data are indistinguishable from
a point source and so hereafter, we assume that all sources are point sources.
The PSF was measured from the real data and normalized by comparing the flux in an 18$\arcsec$ radius
aperture of the PSF with an equivalent SIRTF/TINYTIM (Krist, personal communication) generated PSF with the same FWHM.
Point sources were then added on to the pure noise image with three different
model flux distributions corresponding to the KRR, CE and LDP models. Figure 2 illustrates the difference in source
density between the data and the models. 
Allowing for a matching radius of 2.9$\arcsec$, we find that our SExtractor
parameters result in $\lesssim$1\% of spurious sources in our simulated images where we know the input catalog of sources.
The flux of the extracted source
was measured in a circular aperture of radius 6$\arcsec$ and corrected upwards by a factor 1.8 
to account for the wings of the PSF performed.
We found that as a result of the large source density and the 4.7$\arcsec$ FWHM PSF, using larger
beams results in substantial contamination of the source flux from neighboring sources.

Completeness corrections and flux calibration
were measured using a thorough Monte-Carlo approach. In each iteration of the process, 14 artificial
sources were added to the original data image, and SExtractor run on the resultant new image, with detection and flux
measurement parameters identical to hose used for the original catalogs. The number
of sources added was deliberately kept small to minimize source crowding in the images.
Sources present in the original image were tagged in the new image depending on the change
in their position and/or change in their flux density. Thus, if a source in the new image
had a position within 2.9$\arcsec$ in the artificial image and had a flux difference of less
than 5\% it was tagged as unaffected by the artificial sources. An artifical source
was regarded as detected if one of the untagged sources matched the input position of the artificial source
to within 2.9$\arcsec$.
Detection and flux measurement parameters were identical to those described above.
The process was repeated a 1000 times with
the 14 sources randomly generated from a flat flux distribution between 20$\mu$Jy and 2000$\mu$Jy.
A matrix P$_{ij}$ for the output flux distribution of the artificial sources was generated where $i$ is the
input flux and $j$ the output flux \citep[e.g.][]{sma95}. The nature of the P$_{ij}$ matrix is such that for a particular $i$, the sum
over all $j$ is less than unity. This is the completeness correction factor which was measured to be 
50\% at an input flux value of 35$\mu$Jy.
The observed catalog of sources in the real image was then distributed among the flux bins. The P$_{ij}$ matrix
was renormalized such that the sum over $i$ for a particular $j$ was equal to the number of
detected output sources in that flux bin. The completeness corrected counts in each flux bin $i$ is then the sum over $j$ of the 
renormalized P$_{ij}$ matrix. This approach was tested on the three simulated images, each with their own P$_{ij}$ matrix,
and the input and output catalog compared to ensure the accuracy of the approach.

Uncertainties in the observed counts were assumed to be Poissonian i.e. $\sqrt {\rm N}_{j}$ where N$_{j}$ is the number of 
observed counts in the F$_{j}$ flux bin. These were then propagated through the P$_{ij}$ matrix to derive the uncertainty in the 
completeness corrected counts. Attempts to quantify the systematics in the counts were undertaken. These were dominated by the
adopted size of the beam used to measure the flux and the resultant correction. A 5\% flux calibration error was also introduced
to assess the effect on the counts. The resultant systematic uncertainties are marked as the hatched region on the top panel of Figure 3.

To assess the accuracy of the source extraction at the faintest flux bins, where the completeness fraction is $\sim$30\%,
a pure noise image was generated on a 1.2$\arcsec$ pixel grid
and point sources added with a flux distribution that is identical to the completeness corrected counts at S$_{24\mu m}>20\mu$Jy
but with an extrapolation
to flux densities of 1$\mu$Jy using the CE model. Artificial sources were
added and extracted
from this simulated image using the technique described earlier. The completeness corrected counts in this simulated image are 
only marginally different from those derived from the real data and 
hence we conclude that despite the large completeness corrections, the
shape of the counts at the faint end are reliably determined.

\section{Galaxy Counts and Extragalactic Background Light}

Table 1 and Figure 3 shows the completeness corrected 24 \um\ counts in ELAIS-N1 and the actual number of detected
sources in each flux bin. Also shown in Figure 3 are the model 24\um\ counts from XU, CE, LDP and KRR. All the model counts
are too high at bright flux densities (S$_{24\mu{\rm m}}>500\mu$Jy) with the CE model being the most deviant. 
This could be attributed to cosmic variance and poor statistics on bright objects in this field. 
However, based on the \sst\ 24\um\ counts presented by \citet{pap04}
in this issue which offer much better statistics at these brighter flux densities, it appears that this is a real feature.
This is most likely because the evolutionary models of the various groups were fit to the 
published ISOCAM 15\um\ counts from various different fields 
at flux densities of $\sim$1 mJy which had large associated uncertainties and
have since been shown to be too high \citep[see e.g.][]{grup02, elb99}. 
At flux densities fainter than 400 $\mu$Jy, our counts indicate that both the LDP model 
and KRR model are inconsistent with the observed counts. 
The LDP model is about 30\% too low and has a faint end slope of S$^{-1.7}$ while the KRR model is comparably
low and has a faint end slope of S$^{-1.95}$. 
For comparison, the observed counts in this field at flux densities in the range 20$-$200$\mu$Jy can be fit
by a polynomial of the form dN/dS (arcmin$^{-2}$~$\mu$Jy$^{-1}$)= 92$\times$(S~$\mu$Jy)$^{-1.6\pm0.10}$.
The XU and CE models are within the uncertainties in the counts in this flux range and have S$^{-1.6}$ slopes as well.

To quantify the quality of the model source counts fits to the data at the faint end, we 
performed a minimum sum of absolute errors, a minimum chi-square as well as a 
maximum likelihood analysis of the observed completeness corrected counts since the uncertainty is
dominated by Poissonian errors. 
The likelihood, L is calculated
as: 
\begin{equation}
\ln L = \sum_{i=0}^{i=n}m_{i}\ln(c_{i}) - \sum_{i=0}^{i=n}c_{i}
\end{equation}
The third term in the calculation of the likelihood, $-\Sigma\ln(m_{i}!)$ is irrelevant and is omitted in Equation 1.
$m_{i}$ is the model counts in each of $n$ flux bins and $c_{i}$ is the observed counts corrected for completeness
in those flux bins. We find that for both the minimum absolute errors and chi-square technique, the CE model which includes
both density and luminosity evolution provides 
the best fit to the counts at flux values fainter 500 $\mu$Jy. However, the maximum likelihood technique indicates that the
XU model is marginally a better fit. Thus, we conclude that the CE and XU models both provide reasonably good fits to the
observed data at the faint end. 
Although neither of these models provide good fits at S$_{24\mu{\rm m}}>500\mu$Jy, the counts
at these flux limits are dominated by galaxies at \z$\lesssim$0.5. Roughly speaking, the implication of this poor fit is that
at these redshifts, the LIG component is evolving much more rapidly than the ULIG component.
However, the paucity of sources at faint 24$\mu$m fluxes implies that the rapid increase in the number density of infrared
luminous galaxies 
between redshifts of 0 and 1$-$1.5 that is required by the models does not extend to higher redshifts
and the evolution either reaches a plateau or declines.
Since the optical/ultraviolet luminosity density (with no dust corrections)
also rises but at a slower rate between \z\ab0 and \z\ab1 \citep[see][and references therein]{st99},
the implication is that extinction-corrections to UV-derived star-formation
rate measurements are increasing with redshift, particularly between 0$<z<1$. 

The predicted redshift distribution of the sources (Figure 4) in this flux range as derived from the models
shows a bi-modal structure, peaking at redshift 1 and redshift 2
with $\sim$20\% of sources being at redshifts between 2$<$z$<$3. The bi-modal nature of the redshift distribution is a result of the
two strongest set of PAH features, those around 8$\mu$m and 12$\mu$m, being redshifted into the 24$\mu$m passband.
Luminous infrared galaxies account for about 60\% of 
all sources while ULIGs are about 10\%. The remainder consists of M82-analogs, i.e.
low-redshift objects with L(8-1000\um)$<10^{11}$~L$_{\sun}$. 

The robustness of this redshift distribution can be questioned since it depends on empirical
evolutionary models which broadly fit other multi-wavelength counts but provide poor fits at 
bright 24$\mu$m flux densities.
CE emphasized the degeneracy in evolutionary parameters for the infrared luminosity function i.e. a number
of models which provide reasonable fits to the observed counts can yield varying counts/star-formation rate estimates. 
However, the key conclusion was that all models required a flattening in their (1+z)$^\alpha$ evolution at z$\sim$1.
This is confirmed by the \sst\ observations.
We have re-derived the evolutionary parameters to fit the 24$\mu$m 
counts in this paper and \citet{pap04} as well as the far-infrared
counts in \citet{dol04}. We find that the key variable is the ratio of LIGs to ULIGs which we now evolve independently
unlike in CE. Using the new evolutionary parameters which we will discuss in a later paper, we find that although the fractional
contribution of LIGs and ULIGs can differ slightly from that presented in Figure 4, the 
shape of the overall redshift distribution remains the same.

The CE models derive the total 24\um\ EBL to be 3.7\nwu of which 32\% comes from sources brighter than 1000$\mu$Jy. 
Since the CE model is a factor of $\sim$4 too high at the bright flux densities, the 24$\mu$m EBL must be significantly
lower than this estimate. The 24\um\ IGL from the detected sources in the flux density range 20$<$S$_{\nu}<1000$~$\mu$Jy 
in these \sst\ observations is 1.8$\pm$0.2~\nwu\ while the completeness corrected counts yield a value of 2.0$\pm$0.2~\nwu. 
Thus, we conclude that
these deep \sst\ observations have resolved out at least 50\% of the 24\um\ EBL and probably as much as 65\%. 

\section{Near-Infrared Counterparts of 24$\micron$ Sources}

ISOCAM 15\um\ sources appear to be disparate from the SCUBA 850\um\ sources in that they have relatively bright optical/near-infrared
counterparts which span the range of Hubble-types in morphology. This seems to suggest that they are typically massive galaxies with
stellar mass $\gtrsim10^{10}$~M$_{\sun}$ \citep{fran02}. In contrast, the SCUBA 850\um\ sources are optically faint
and mostly irregular systems \citep{chap02}. This could be a result of morphological $k-$correction since visible light observations
of SCUBA sources trace rest-frame ultraviolet emission which could be patchy due to dust extinction internal to the galaxy. 
Since the models suggest that the 24\um\ sources presented here are at redshifts
intermediate to the ISOCAM and 850\um\ sources, this would imply that they have rest-frame properties that 
are intermediate to those sources.

We matched sources between MIPS and IRAC using a nearest neighbor
match between the 24\um\ catalog and the 3.6 and 4.5 \um\ catalogs (Figure 5).  At a 24\um\ flux
limit of 50$\mu$Jy, more than 85\% of the sources have IRAC counterparts
within 3$\arcsec$, as compared to the 25\% expected for random matches to objects with
the same density; thus at least 60\% of the matches are physical.

Direct inspection of the remaining 15\% shows that in almost all cases, separations
of 3$\arcsec$ or larger arise not from a lack of physical counterpart, 
but due to a variety of other factors: 7.8\% of MIPS sources fall in a region of low exposure time in either the IRAC image or MIPS image
or both;
3.6\% are cases where the IRAC counterpart was not automatically detected because it falls near a much
brighter IRAC sources and was not deblended from the brighter IRAC sources;
6.5\% of MIPS sources appear to be blended emission from two or more nearby IRAC sources in which
case the MIPS source appears slightly extended/distorted and its centroid falls between the IRAC positions which are more
than 3$\arcsec$ away.
There are however 0.9\% of cases where the MIPS source genuinely appears to have no IRAC counterpart to the sensitivity
of the IRAC image. Since the distribution of MIPS/IRAC flux ratios is broad, depending on the nature of the source
and its redshift, faint MIPS sources could 
have counterparts which are below the flux threshold of the IRAC observations. 
Therefore we conclude that the fraction of 24\um\ sources with reliable near-infrared associations
is $\sim$100\%.

Having identified the counterparts of the sources, we then proceed to estimate a flux ratio between the mid-infrared and
near-infrared wavelengths. 
For the MIPS catalog we adopted a beam size corrected flux value as described in Section 2. For 
the near-infrared we adopted SExtractor flux values in a 6$\arcsec$ diameter beam and applied a 20\% beamsize correction based
on estimates published by the \sst\ Science Center.
Figure 5 illustrates the mid-infrared to near-infrared flux ratios of the Spitzer 24\um\ sources detected here. 
Also shown as solid circles is the 15\um\ to HK-band flux ratio \citep[partly from][and partly from the GOODS data]{bar02}
for the ISOCAM 15\um\ sources in the HDF-N which have spectroscopically determined
redshifts peaking at $\sim$0.8. Due to the shape of the dust spectral energy distribution at these wavelengths, sources 
at the same redshift
will mostly be brighter at observed 24\um\ than at observed 15\um. However, at $z\sim2$, the properties of 24\um\ sources
should trace the same rest-frame emission as the 15\um\ observations at z\ab1. 
Since, the mid-infrared flux of a galaxy traces its total star-formation rate, while the near-infrared flux
is a measure of the stellar mass in the galaxies, the similarity in this flux
ratio between the ISOCAM and MIPS sources broadly suggests
that the specific star-formation rate i.e. the ratio of SFR to stellar mass, is very similar among the \z\ab 1 ISOCAM galaxies
and \z\ab 1-2 MIPS galaxies.

\acknowledgements
Support for this work, part of the Spitzer Space Telescope Legacy Science Program, was provided by NASA through an award
issued to the Jet Propulsion Laboratory, California Institute of Technology under NASA contract 1407. We wish to acknowledge the \sst\ Science
Center director, Tom Soifer for allocating Director's Discretionary Time for these observations. We also thank the
referee for useful comments.

\clearpage

\begin{figure}
\plotone{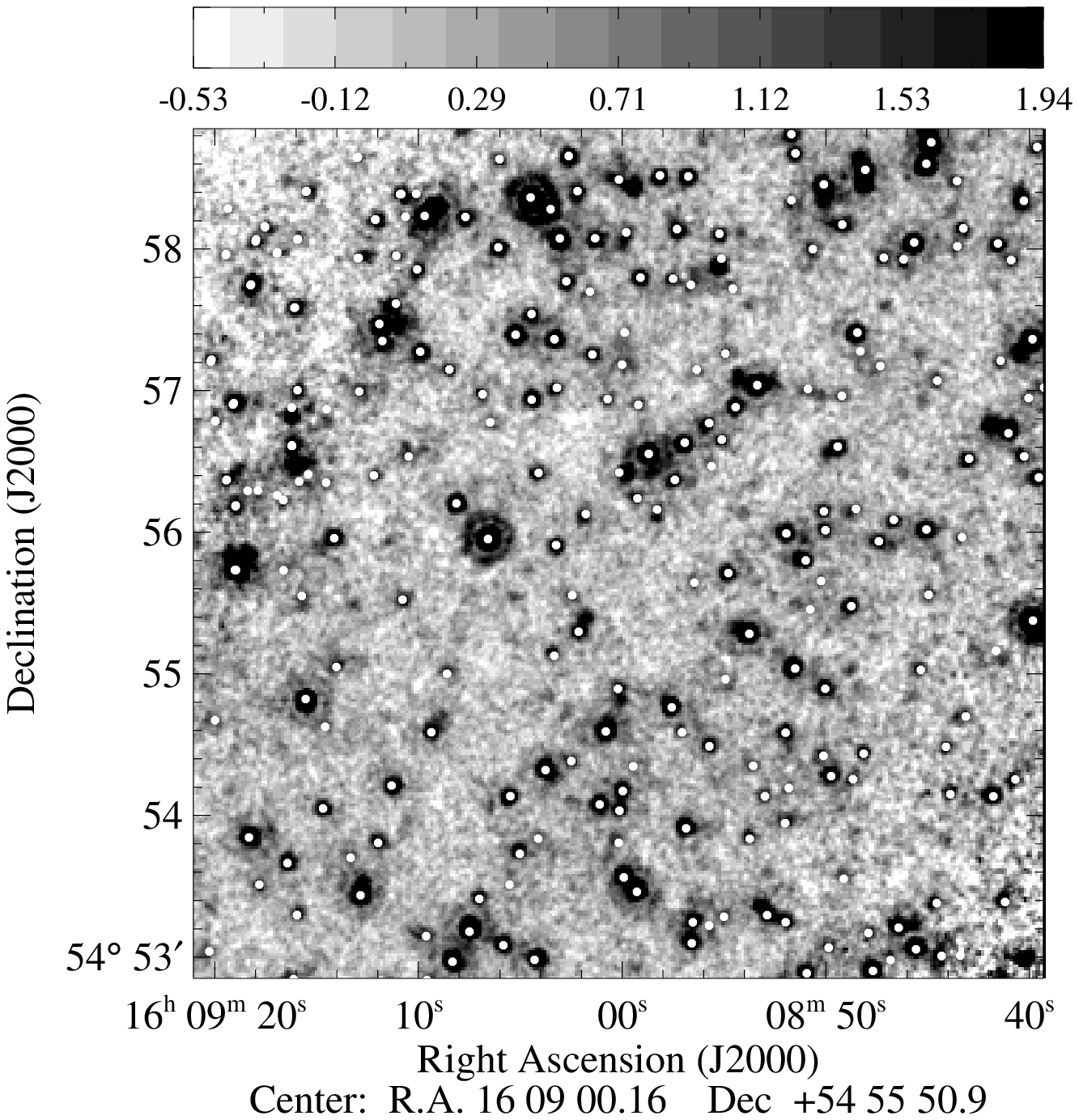}
\caption{
Image showing the quality of the MIPS 24\um\ data in the ELAIS-N1 field. The stretch is in flux values in units of microJy.
The dots indicate the position of the IRAC counterpart. 
}
\end{figure}

\begin{figure}
\epsscale{0.6}
\plotone{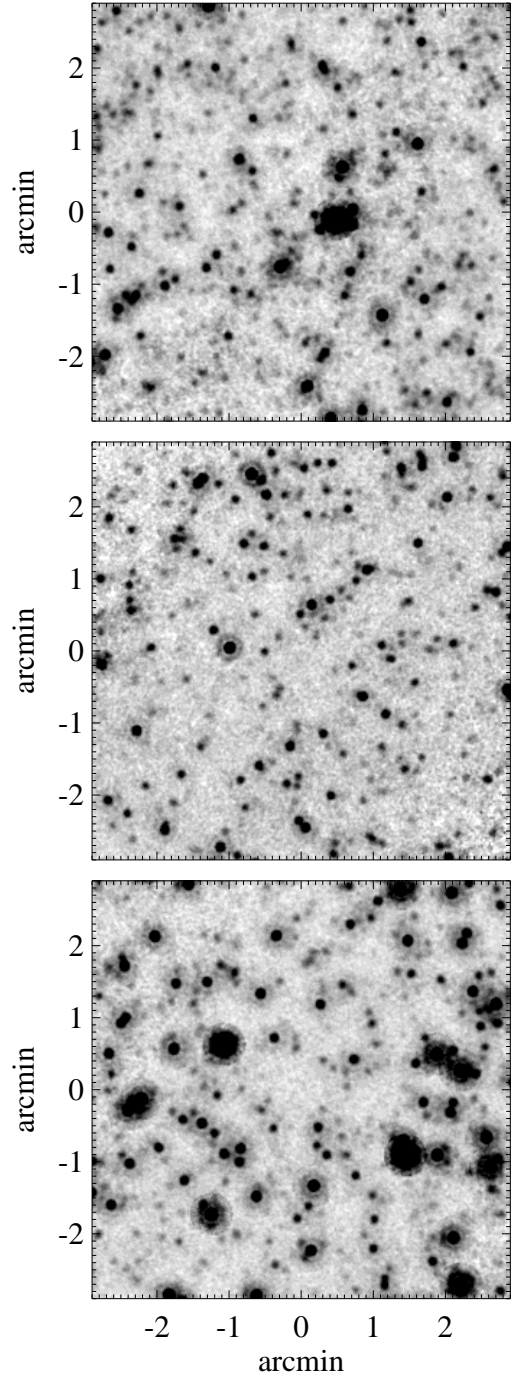}
\caption{
Comparison between the source densities in the real data (center) and the simulated image from the King \& Rowan-Robinson model (top)
and the Chary \& Elbaz model (bottom). Clearly the KRR model predicts too many faint sources while the CE model has too many bright
sources with respect to the data.
}
\end{figure}

\begin{figure}
\epsscale{1.0}
\plotone{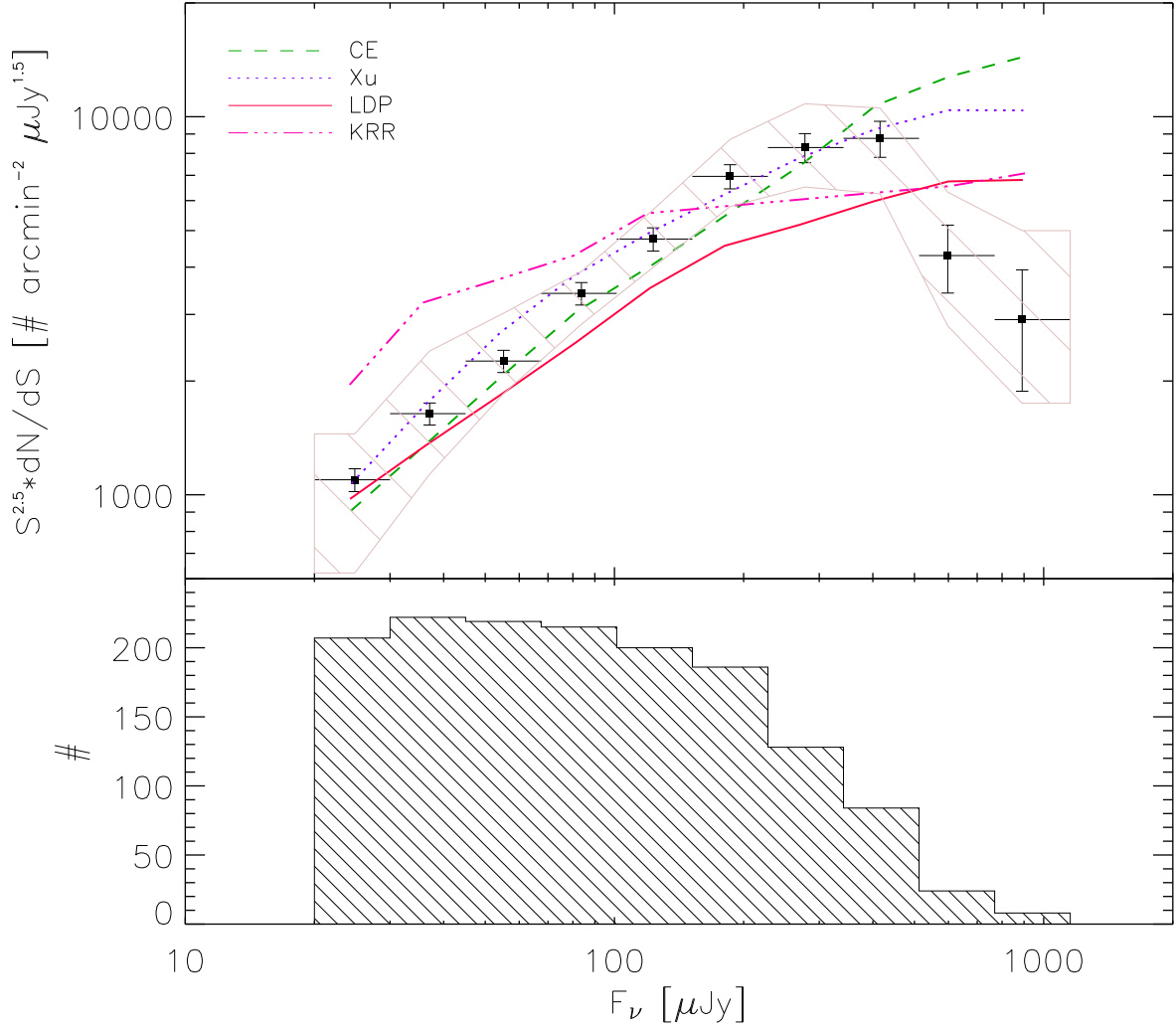}
\caption{Completeness corrected galaxy counts in the MIPS 24\um\ channel
from \sst\ observations of ELAIS-N1. The error bars reflect the Poissonian uncertainty. 
The horizontal bars represent the minimum and maximum flux
density in that bin. 
The lines show four preferred models for 24\um\ counts (see text). 
The symbols are plotted at the average of the flux densities
of the detected sources in that bin for the data while the lines are plotted at the counts-weighted flux average
for the models. The lower plot in the figure shows the histogram of the actual number of sources detected in each flux bin without
any completeness correction.
}
\end{figure}

\begin{figure}
\plotone{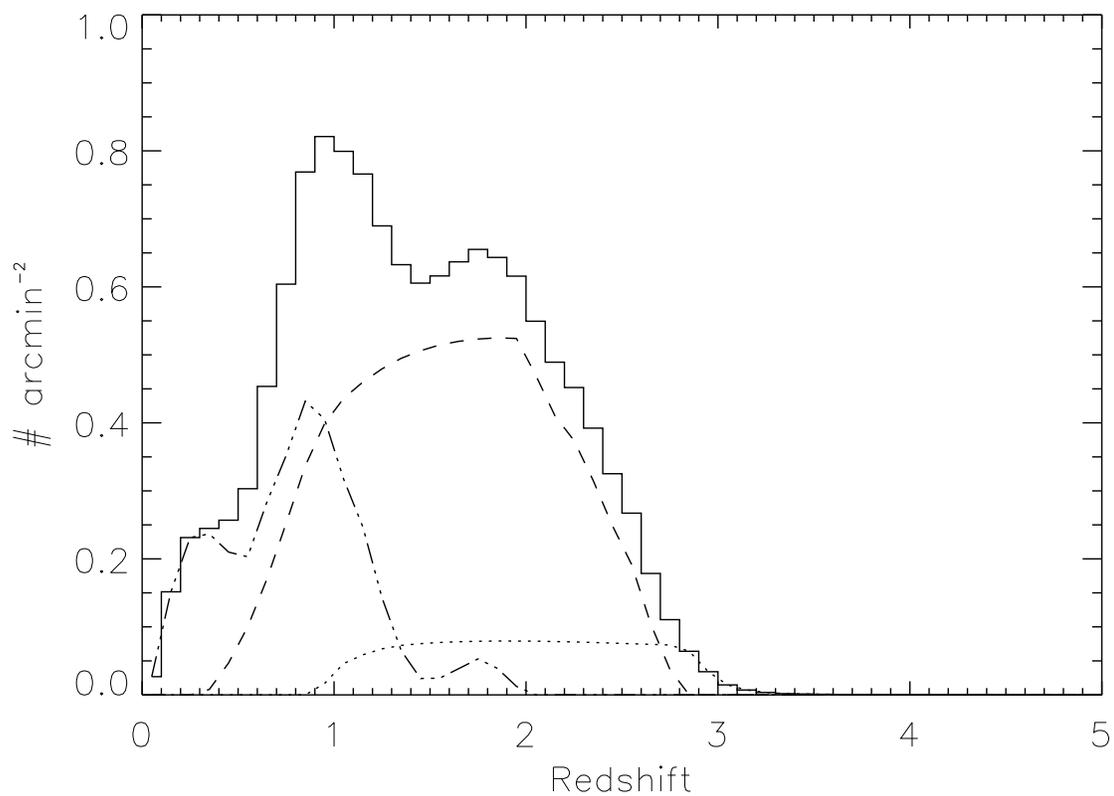}
\caption{Redshift distribution expected for 24\um\ sources in the flux range 20$<S_{\nu}<500\mu$Jy as derived
from the best fitting CE model. The dotted line
indicates the contribution from ULIGs (L$_{\rm IR}=$L(8-1000\um)$>10^{12}$~L$_{\sun}$), 
the dashed line is for LIGs with the triple-dot-dashed line the contribution
from low-luminosity starbursts with L$_{\rm IR}<10^{11}$~L$_{\sun}$. The size of the redshift bins is 0.1. 
}
\end{figure}

\begin{figure}
\plotone{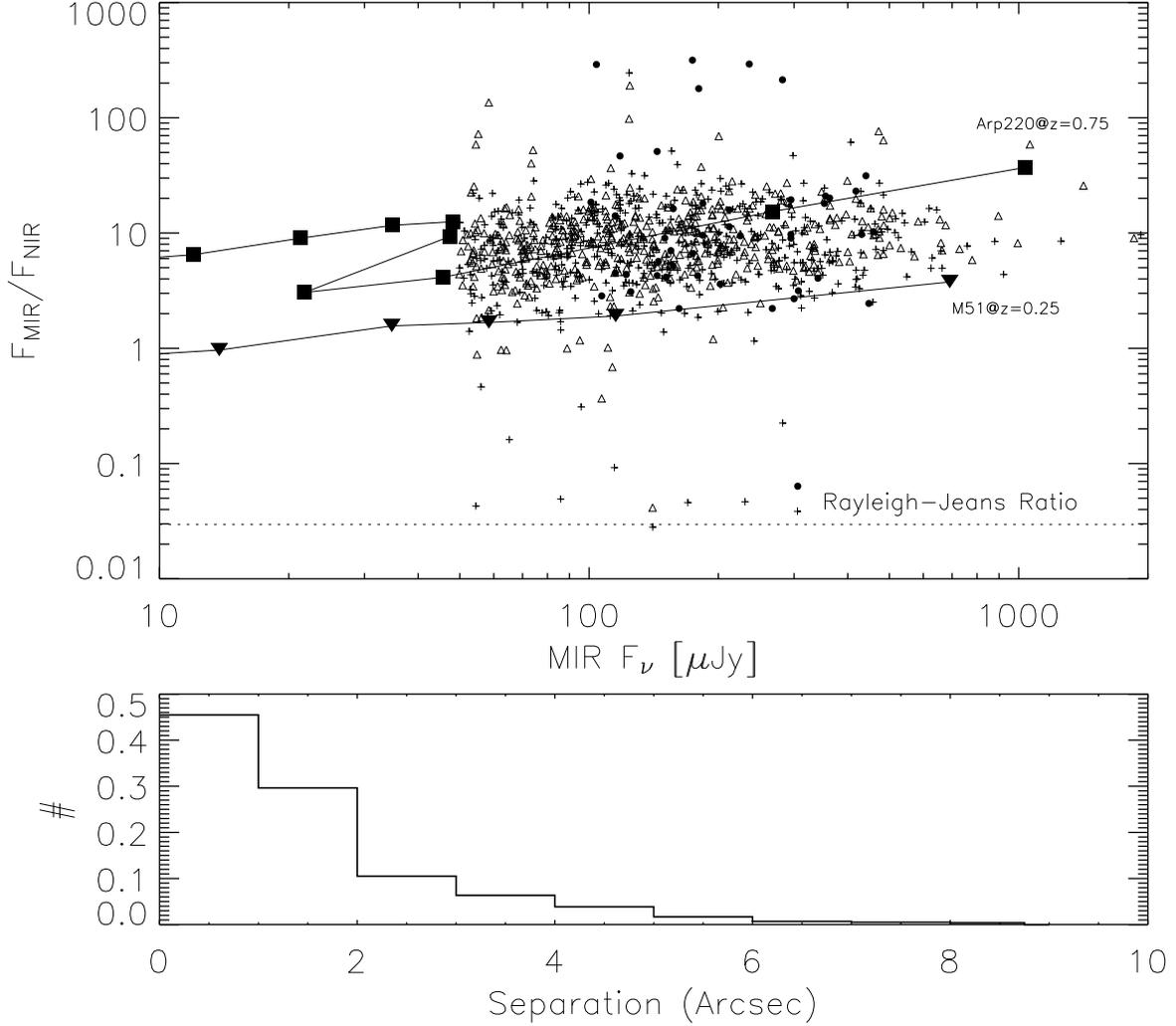}
\caption{Ratio between the mid-infrared (MIR; 24\um) flux density and the near-infrared (NIR)
flux density in the 3.6\um\ (plus) or 4.5\um\ (triangles) IRAC channels for all 24\um\ sources brighter than 50$\mu$Jy. 
Also shown in the lower
plot is the distribution of separation in arcseconds between the MIPS source and its IRAC counterpart. 
Of the \ab1000 sources brighter than 50$\mu$Jy in our 24\um\
catalog, more than 85\% have associations in the IRAC images which are less than 3$\arcsec$ away. 
The plot also shows the 15\um\ to HK-band flux ratio for the
ISOCAM sources in HDF-N (solid circles). 
To qualify the nature of the MIPS sources, we also show the evolution of the colors of two 
nearby galaxies; the ultraluminous starburst Arp220 (solid squares)
and the quiescent starburst/spiral galaxy M51 (solid triangle) as a function of redshift.
The redshift increases in steps of 0.25 with the redshift of the first point as labelled. Arp220 and M51 at lower redshift will be to
the right of this plot.
}
\end{figure}

\begin{deluxetable}{cccccc}
\tabletypesize{\scriptsize}
\tablecaption{24\um\ Counts in the ELAIS-N1 field\tablenotemark{a}}
\tablewidth{0pt}
\tablehead{
\colhead{S$_{low}$} & \colhead{S$_{high}$} & \colhead{Avg S$_{\nu}$} & \colhead{Observed dN/dS} & \colhead{Corrected dN/dS} &
\colhead{Uncertainty dN/dS}
}
\startdata
  20.0&  30.0&  24.8& 1.2E-01& 3.6E-01& 2.5E-02\\
  30.0&  45.0&  37.1& 8.9E-02& 2.0E-01& 1.3E-02\\
  45.0&  67.5&  55.2& 5.9E-02& 1.0E-01& 6.8E-03\\
  67.5& 101.2&  83.8& 3.8E-02& 5.3E-02& 3.6E-03\\
 101.2& 151.9& 123.1& 2.4E-02& 2.8E-02& 2.0E-03\\
 151.9& 227.8& 185.9& 1.5E-02& 1.5E-02& 1.1E-03\\
 227.8& 341.7& 277.9& 6.8E-03& 6.4E-03& 5.7E-04\\
 341.7& 512.6& 415.8& 3.0E-03& 2.5E-03& 2.7E-04\\
 512.6& 768.9& 597.9& 5.6E-04& 4.9E-04& 1.0E-04\\
 768.9&1153.3& 891.2& 1.3E-04& 1.2E-04& 4.3E-05\\
\enddata
\tablenotetext{a}{S$_{\nu}$ is in $\mu$Jy while dN/dS is in \#~arcmin$^{-2}$~$\mu$Jy$^{-1}$. Uncertainty in dN/dS is the Poissonian
noise propagated through the P$_{ij}$ matrix.}

\end{deluxetable}

\end{document}